\documentclass[12pt]{article}

\usepackage{amsmath,amssymb,graphicx,multicol} 
\usepackage{epsf,color}
\usepackage{pstricks}
\usepackage{cite}
\usepackage{verbatim}
\usepackage{slashed}

\newcommand{\beq}{\begin{eqnarray}}
\newcommand{\eeq}{\end{eqnarray}}

\newcommand{\centeron}[2]{{\setbox0=\hbox{#1}\setbox1=\hbox{#2}\ifdim
                                        
\wd1>\wd0\kern.5\wd1\kern-.5\wd0\fi
\copy0

\kern-.5\wd0\kern-.5\wd1\copy1\ifdim\wd0>\wd1
                                       \kern.5\wd0\kern-.5\wd1\fi}}
\newcommand{\ltap}{\>\centeron{\raise.35ex\hbox{$<$}}
                               {\lower.65ex\hbox{$\sim$}}\>}
\newcommand{\gtap}{\>\centeron{\raise.35ex\hbox{$>$}}
                               {\lower.65ex\hbox{$\sim$}}\>}

\newcommand\ZZ{\hbox{\zfont Z\kern-.4emZ}}
\font\zfont = cmss10 

\begin{document}
\begin{titlepage}
\vskip .5cm
\begin{center}
{\huge \bf Constraints on the Unhiggs Model}
\vskip .5 cm
{\huge \bf  from Top Quark Decay}

\vskip .1cm
\end{center}
\vskip .2cm

\begin{center}
{\bf
{David Stancato}, {\rm and}
{John Terning}}
\end{center}
\vskip 8pt

\begin{center}
{\it
Department of Physics, University of California, Davis, CA  
95616} \\
\vspace*{.5cm}
\vspace*{0.3cm}
{\tt  dastancato@ucdavis.edu, jterning@gmail.com}
\end{center}

\vglue 0.3truecm

\begin{abstract}
\vskip 3pt
\noindent
We compute the top quark decay rate in the Unhiggs model. In this model, the longitudinally polarized $W$'s are unparticles, which is owed to their Goldstone boson nature, while the transversely polarized $W$'s are not. Thus the fraction of decays with a longitudinal $W$ emitted is different than in the Standard Model. Comparing this calculation to CDF data, we are able to rule out some of the Unhiggs model parameter space. We also use the expected increased accuracy of top decay measurements at the LHC to anticipate further constraints on the Unhiggs.

\end{abstract}

\end{titlepage}

\newpage

\setcounter{footnote}{0}

\section{Introduction}
\label{sec:intro}
\setcounter{equation}{0}

Georgi \cite{Georgi, Georgi2} introduced a new approach to studying conformal sectors by specifying the two-point functions of fields with a scaling dimension between one and two. Since the phase space for these fields resembles the phase space of a fractional number of particles, Georgi termed them ``unparticles''. Subsequently, efforts were made to gauge unparticle actions in a consistent way \cite{coloredunparticles} so that unparticles could be given Standard Model (SM) gauge quantum numbers. This also necessitates the introduction of an IR cutoff to the unparticle sector \cite{Irvine,coloredunparticles,AdSCFTUnP} so that there are no new massless modes, which would dramatically alter low energy phenomenology. In a previous paper \cite{Unhiggs}, we introduced the Unhiggs as a way to break electroweak symmetry via an unparticle (see ref. \cite{mixing,other} for work on related ideas). The Unhiggs has the same gauge structure as the SM Higgs and has an IR cutoff $\mu$. The effects of an Unhiggs on precision electroweak measurements have been studied in detail \cite{Falkowski}, and the model is consistent with the current data. In \cite{Unhiggs}, we showed that like the SM Higgs, the Unhiggs unitarizes $WW$ scattering. The Unhiggs also has some advantages over the SM Higgs. Because the SM Higgs has scaling dimension one, its mass is quadratically sensitive to the scale of new physics. The fact that electroweak precision tests prefer a low Higgs mass compared to the TeV scale thus seems somewhat fine-tuned and is sometimes known as the Little Hierarchy Problem. The Unhiggs, however, has a scaling dimension greater than one and should thus be less sensitive to the scale of new physics. This was borne out explicitly in \cite{Unhiggs}, as we showed that the scale of new physics could be pushed above a TeV without much fine-tuning, thus ameliorating the Little Hierarchy Problem. 

In this paper, we would like to begin investigating possible phenomenological bounds on the Unhiggs model. We are motivated in this direction by a result from Georgi's first unparticle paper \cite{Georgi}, in which he computes the decay rate of the top quark in a toy model consisting of the top quark, the up quark and a generic unparticle scalar field with scaling dimension $d$. He found that when the scalar field is an unparticle with $1 < d < 2$, the decay rate differs, in some cases dramatically, from the decay rate when the scalar field is a standard particle with $d=1$. In the Unhiggs model, we expect to see something similar because of the fact that both the physical Unhiggs and the Goldstone bosons are unparticles. Since the Goldstone bosons are ``eaten'' by the $W^{\pm}$ and $Z$ gauge bosons, the longitudinal components of $W^{\pm}$ and $Z$ will also exhibit unparticle behavior. This can be seen explicitly in the form of the gauge boson propagators derived in \cite{Unhiggs}. Thus, we can use top decay, $t \rightarrow W^+ b$, to investigate some phenomenological consequences of the Unhiggs model. Since only the longitudinal component of the $W^+$ is an unparticle, while the transverse components are just standard gauge bosons, we expect to find that the fraction of top decays with a longitudinally produced $W$ boson will depend on $d$, and thus will generically differ from the result in the Standard Model (SM). We also expect the fraction to depend on the IR cutoff, or threshold, $\mu$, which is an is an important phenomenological parameter in the Unhiggs construction. This is due to the fact that it serves to cut off the low energy part of the Unhiggs continuum; the value of $\mu$ corresponds to the energy scale at which the Unhiggs continuum begins. Without this IR cutoff there would be no mass gap and the continuum of states corresponding to the Unhiggs would introduce modes with very low mass, which obviously contradicts experiment. The fact that the threshold must be large enough to avoid this problem will manifest itself in the constraints that our calculation will place on $\mu$. To place constraints on the $\mu$-$d$ parameter space, we will compare the fraction of top decays with longitudinally produced $W$ bosons as calculated in the Unhiggs model with the experimental data from CDF. Finally, we use the expected increased sensitivity of this measurement at the LHC to obtain future expected bounds on the Unhiggs model.  

\section{Calculation of the top decay rate}
\label{sec:calcrate}
\setcounter{equation}{0}

Since we want to find the fraction of decays with a longitudinally produced $W$, we separately calculate the decay rates for transversely and longitudinally emitted $W$ bosons. Since the transverse gauge bosons are unaffected by the electroweak symmetry breaking sector, the top decay rate to transverse $W$'s, $\Gamma (t\rightarrow W_T^+ b)$, will remain the same as in the Standard Model. This decay rate is given by
\begin{align}
\label{eqn:GammaTransW}
\Gamma(t\rightarrow W^+_T b) = \frac{g^2}{32\pi}m_t\left(1-\frac{M_W^2}{m_t^2}\right)^2 ~.
\end{align}

Since longitudinal gauge bosons are intimately connected to the electroweak symmetry breaking sector, we expect that the unparticle Goldstone bosons of the Unhiggs model will have an effect on the decay rate for longitudinally produced $W$ bosons. We will indeed find that this decay rate differs from its value in the SM. 
The physical decay rate is the sum of the decay rate to a longitudinal $W$ and the decay rate to a Goldstone boson in a generic $R_\xi$ gauge.
\begin{align}
\Gamma_{GI}(t\rightarrow W^+_L b) = \Gamma_{R_\xi}(t\rightarrow W^+_L b) + \Gamma_{R_\xi}(t\rightarrow \pi^+ b)
\end{align}
where $\Gamma_{GI}(t\rightarrow W^+_L b)$ is the gauge invariant, physical decay rate, and $\Gamma_{R_\xi}(t\rightarrow W^+_L b)$ and $\Gamma_{R_\xi}(t\rightarrow \pi^+ b)$ are the gauge dependent decay rates to $W_L^+$ and $\pi^+$, respectively. 
To begin, we need the squared amplitudes for $W_L^+$ production as well as for $\pi^+$ production. Since the Unhiggs does not affect the fermion-gauge boson couplings, $|\mathcal{M}_{W_L^+}|^2$ will be the same as in the SM and is given by
\begin{align}
\label{eqn:WAmp}
|\mathcal{M}_{W_L^+}|^2 &= \frac{g^2}{2}(q^\mu p^\nu + p^\mu q^\nu -g^{\mu \nu}q\cdot p)\epsilon_{\mu}(k) \epsilon_{\nu}(k) \\ \nonumber
& = \frac{g^2}{2}\frac{m_t}{M_W^2}|\vec{q}|\left(k^0+|\vec{k}|\right)^2~,
\end{align}
where $p$ is the 4-momentum of the top quark, $q$ is the 4-momentum of the bottom quark and $k$ is the 4-momentum of the $W_L^+$ boson. To find $|\mathcal{M}_{\pi^+}|^2$ we must make use of the fact that the Unhiggs effective Lagrangian contains a Yukawa coupling term of the form
\beq
\label{eqn:Yukawa}
\mathcal{L} \ni -\lambda_t \bar{t}_{R} \frac{H^{\dagger}}{\Lambda^{d-1}} \left( \begin{array}{c} t \\ b \end{array} \right) + \textrm{ h.c.}~,
\eeq
where $H$ is the Unhiggs doublet, $\lambda_t$ is the top Yukawa coupling, $\Lambda$ is the UV cutoff of the effective theory and $d$ is the Unhiggs scaling dimension which is restricted to the range $1 \leq d < 2$ . We also note the following relations which are derived in \cite{Unhiggs}:
\begin{eqnarray}
m_t &=& \frac{\lambda_t v^{d}}{\sqrt{2}\Lambda^{d-1}}~, \\ \nonumber
M_W^2 &=& \frac{g^2 (2-d)\mu^{2-2d}v^{2d}}{4}~,
\end{eqnarray}
where $v^{d}$ is the Unhiggs VEV and $\mu$ is the Unhiggs threshold mass. Using these relations along with the vertex derived from Equation \ref{eqn:Yukawa}, we find that
\beq
\label{eqn:PiAmp}
|\mathcal{M}_{\pi^+}|^2 = \frac{g^2}{2}\frac{m_t^3}{M_W^2}(2-d)\mu^{2-2d}|\vec{q}|~,
\eeq 
where $q$ is the 4-momentum of the bottom quark.

The non-trivial part of the decay rate calculation is determining the various phase space factors. To accomplish this, we will need to make use of the following results from \cite{Unhiggs}:
The $W_L^{\pm}$ propagator is given by
\begin{align}
\Delta_{W_L^{\pm}}(k) = \frac{-i}{k^2-M_W^2}\left(\frac{\xi \left(k^2 -  M_W^2\right)\mu^{2-2d}-f(k^2)\left(1-\frac{\xi  \,M_W^2}{(2-d)k^2}\right)}{f(k^2)\left(k^2-\frac{\xi M_W^2}{2-d}\right)} k_\alpha k_\beta \right)
\end{align}
where  
\begin{align}
f(k^2) \equiv \mu^{4-2d}-(\mu^2-k^2)^{2-d}~.
\end{align}
In addition, the Goldstone boson propagator is given by
\begin{align}
\Delta_{\pi^\pm}(k)&=
\frac{i}{f(k^2) +\xi \frac{M_W^2}{2-d}f(k^2)/k^2}~.
\end{align}
Calculating the phase space factors will be easiest in Landau gauge ($\xi = 0$) because the longitudinal $W_L^{\pm}$ propagator reduces to that of the corresponding SM Landau gauge propagator, which is given by
\begin{align}
\Delta_{W_L^{\pm},\xi =0}(k) = \frac{i}{k^2-M_W^2}\frac{k^\mu k^\nu}{k^2} = \frac{k^\mu k^\nu}{M_W^2}\left(\frac{i}{k^2-M_W^2}-\frac{i}{k^2}\right)~.
\end{align}
Since $\Delta_{W_L^{\pm},\xi=0}(k)$ has two simple poles at $k^2=0$ and $k^2=M_W^2$, the phase space takes the following simple form in this gauge:
\beq
\label{eqn:LanWphase}
d\Phi_{W_L^{\pm},\xi=0}(k) = 2\pi \theta(k^0) \delta(k^2-M_W^2) -2\pi \theta(k^0) \delta(k^2)~.
\eeq
The Goldstone boson propagator in Landau gauge is given by
\begin{align}
\label{eqn:LanPiProp}
\Delta_{\pi^{\pm},\xi=0}(k) = \frac{i}{f(k^2)}~.
\end{align}
The phase space in this case is not as simple as for $W_L^{\pm}$. To find its form, we use the Unhiggs propagator and the resulting Unhiggs phase space as derived in \cite{Unhiggs}. The Unhiggs propagator is
\begin{align}
\label{eqn:Unhprop}
\Delta_h(k^2) = \frac{i}{f(k^2)-m^{4-2d}}
\end{align}
while the Unhiggs phase space is
\beq \nonumber
\label{eqn:Unhphase}
d\Phi_h(k^2) &=& \frac{-2\sin(\pi d)\theta(k^0)\theta(k^2 - \mu^2) (k^2-\mu^2)^{2-d}}{(\mu^{4-2d}-m^{4-2d})^2 + (k^2-\mu^2)^{4-2d} - 2(\mu^{4-2d}-m^{4-2d})(k^2-\mu^2)^{2-d} \cos (d\pi)} \\
&+& 2\pi \theta(k^0)\frac{(\mu^{4-2d}-m^{4-2d})^{\frac{d-1}{2-d}}}{(2-d)}\delta \left[k^2 - \mu^2 - \left(\mu^{4-2d}-m^{4-2d}\right)^{\frac{1}{2-d}}\right]~.
\eeq
Since the Goldstone boson propagator in Landau gauge, Eq. (\ref{eqn:LanPiProp}), is equal to the Unhiggs propagator with $m=0$, we use the Unhiggs phase space with $m=0$ to find the following form for the Goldstone boson phase space:
\begin{align} 
d\Phi_{\pi^{\pm},\xi=0}(k) =& \frac{-2\sin (\pi d) \theta (k^0) \theta (k^2-\mu^2)(k^2-\mu^2)^{2-d}}{\mu^{8-4d}+(k^2-\mu^2)^{4-2d} -2\mu^{4-2d}(k^2-\mu^2)^{2-d}\cos{d\pi}} \\ \nonumber
& + 2\pi \theta (k^0) \frac{\mu^{2d-2}}{2-d}\delta(k^2)~. 
\end{align}
Note that this phase space is a sum of a pole at $k^2 = 0$ and a continuum above the unparticle threshold $\mu$. It will be useful to separate these two factors and write the Goldstone boson phase space as
\begin{align}
\label{eqn:Piphase2}
d\Phi_{\pi^{\pm},\xi=0}(k) \equiv d\Phi^{(1)}_{\pi^{\pm},\xi=0}(k) + d\Phi^{(2)}_{\pi^{\pm},\xi=0}(k)
\end{align}
where
\begin{align}
d\Phi^{(1)}_{\pi^{\pm},\xi=0}(k) = \frac{-2\sin (\pi d) \theta (k^0) \theta (k^2-\mu^2)(k^2-\mu^2)^{2-d}}{\mu^{8-4d}+(k^2-\mu^2)^{4-2d} -2\mu^{4-2d}(k^2-\mu^2)^{2-d}\cos{d\pi}}
\end{align}
and
\begin{align}
d\Phi^{(2)}_{\pi^{\pm},\xi=0}(k) =  2\pi \theta (k^0) \frac{\mu^{2d-2}}{2-d}\delta(k^2)~.
\end{align}

Armed with the phase space factors, we will now calculate the Landau gauge decay rates to $W_L^+$ and $\pi^+$, respectively. Since the $W^+_L$ phase space is just the sum of two delta function factors, the decay rate for $W_L^+$ emission in Landau gauge is easily found to be
\begin{align}
\Gamma_{W_L^+,\xi=0} = \frac{1}{16\pi m_t^3}(m_t^2-M_W^2)|\mathcal{M}_{W_L^+}(k^2=M_W^2)|^2 - \frac{1}{16\pi m_t}|\mathcal{M}_{W_L^+}(k^2 = 0)|^2~. 
\end{align}
Using Eq. (\ref{eqn:WAmp}), we find
\begin{align}
\label{eqn:LanGammaW}
\Gamma_{W_L^+,\xi=0} = \frac{g^2}{64\pi}\frac{m_t^3}{M_W^2}\left(1-\frac{M_W^2}{m_t^2}\right)^2 - \frac{g^2}{64\pi}\frac{m_t^3}{M_W^2}~.
\end{align}
To find the decay rate for $\pi^+$ emission in Landau gauge, we separate $d\Gamma_{\pi^+,\xi=0}$ into two parts, corresponding to the separation of the phase space factors in Eq. (\ref{eqn:Piphase2}).
\beq
d\Gamma_{\pi^+,\xi=0} \equiv d\Gamma^{(1)}_{\pi^+,\xi=0} + d\Gamma^{(2)}_{\pi^+,\xi=0}
\eeq
where
\begin{align}
d\Gamma^{(1)}_{\pi^+,\xi=0} = \frac{1}{2m_t}d\Phi^{(1)}_{\pi^+,\xi=0}(k)d\Phi_b(q)|\mathcal{M}_{\pi^+}|^2 (2\pi)^4 \delta^4 (p-k-q)~,
\end{align}
and
\begin{align}
d\Gamma^{(2)}_{\pi^+,\xi=0} = \frac{1}{2m_t}d\Phi^{(2)}_{\pi^+,\xi=0}(k)d\Phi_b(q)|\mathcal{M}_{\pi^+}|^2 (2\pi)^4 \delta^4 (p-k-q)~.
\end{align}
Here $k$ is the 4-momentum vector of the $\pi^+$ and $\mathcal{M}_{\pi^+}$ is given by Eq. (\ref{eqn:PiAmp}).
Focusing first on $d\Gamma^{(1)}_{\pi^+,\xi=0}$, we find
\begin{align}
\Gamma_{\pi^+,\xi=0}^{(1)} &= \frac{1}{2m_t}\frac{g^2}{2(2\pi)^3}\frac{m_t^2}{M_W^2}(2-d)\mu^{2-2d}\int d^4k d\Phi^{(1)}_{\pi^+,\xi=0}(k)\frac{m_t}{2} \delta(m_t-k^0-|\vec{k}|) \\ \nonumber
& = \frac{g^2 m_t^2}{64\pi^3 M_W^2}(2-d)\mu^{2-2d}\int d^4k d\Phi^{(1)}_{\pi^+,\xi=0}(k) \delta(m_t-k^0-|\vec{k}|)~.
\end{align}
To simplify the calculation of $d\Gamma^{(1)}_{\pi^+,\xi=0}$, let
\begin{align}
\label{eqn:g}
g(k^2-\mu^2) \equiv \frac{-2\sin (\pi d) (k^2-\mu^2)^{2-d}}{\mu^{8-4d}+(k^2-\mu^2)^{4-2d}-2\mu^{4-2d}(k^2-\mu)^2\cos (d\pi)}~.
\end{align}
Then we have
\begin{align}
\Gamma_{\pi^+,\xi=0}^{(1)} = &\frac{1}{2m_t}\frac{g^2}{2(2\pi)^3}\frac{m_t^2}{M_W^2}(2-d)\mu^{2-2d} \\ \nonumber
&\times \int d^4k \theta(k^0)\theta(k^2-\mu^2)g(k^2-\mu^2)\delta(m_t-k^0-|\vec{k}|)~. 
\end{align}
Next, we write $d^4k = dk^0 d^3k = dk^0 |\vec{k}|^2 d|\vec{k}|d\Omega$, complete the angular integration and then perform the delta function integration over $|\vec{k}|$ to get
\begin{align}
\Gamma_{\pi^+,\xi=0}^{(1)} = &\frac{1}{2m_t}\frac{g^2}{(2\pi)^2}\frac{m_t^2}{M_W^2}(2-d)\mu^{2-2d} \\ \nonumber
&\times \int_0^{m_t} dk^0 (m_t-k^0)^2 \theta(2m_t k^0-m_t^2-\mu^2)g(2m_tk^0-m_t^2-\mu^2)~.
\end{align}
The theta function here simply enforces the fact that the phase space for the unparticle is zero below threshold and therefore that the decay rate is zero unless $m_t \geq \mu$. With the substitution $x = 2m_tk^0 -m_t^2-\mu^2$, we finally obtain the following result:
\begin{align}
\label{eqn:LanGammaPi1}
\Gamma_{\pi^+,\xi=0}^{(1)} = \frac{g^2}{128\pi^2}\frac{m_t^3}{M_W^2}(2-d)\mu^{2-2d}\int_0^{m_t^2-\mu^2} \left(\frac{m_t^2-\mu^2-x}{m_t^2}\right)^2 g(x)dx
\end{align}
with $g(x)$ given by Eq. (\ref{eqn:g}).  
The evaluation of $\Gamma_{\pi^+,\xi=0}^{(2)}$ is much easier since the phase space $d\Phi_{\pi^+,\xi=0}^{(2)}$ is just $\frac{\mu^{2d-2}}{2-d}$ times the phase space for a single particle of mass zero. Therefore we simply have a decay of the top quark into two massless particles. The decay rate for this process is given by
\begin{align}
\Gamma_{\pi^+,\xi=0}^{(2)} = \frac{1}{8\pi}\frac{1}{2m_t}\frac{\mu^{2d-2}}{2-d}|\mathcal{M}_\pi^+|^2~.
\end{align}
Using $|\vec{q}| = \frac{m_t}{2}$ in the massless case, we find
\begin{align}
\label{eqn:LanGammaPi2}
\Gamma_{\pi^+,\xi=0}^{(2)} = \frac{g^2}{64\pi}\frac{m_t^3}{M_W^2}~. 
\end{align}
Combining Eqs. (\ref{eqn:LanGammaW}), (\ref{eqn:LanGammaPi1}) and (\ref{eqn:LanGammaPi2}) we arrive at the gauge invariant decay rate of the top quark with a longitudinal $W$ in the final state:
\begin{align}
\label{eqn:GammaGILongW}
\Gamma_{GI}(t\rightarrow W_L^+ b) =& \frac{g^2}{128\pi^2}\frac{m_t^3}{M_W^2}(2-d)\mu^{2-2d}\int_0^{m_t^2-\mu^2} \left(\frac{m_t^2-\mu^2-x}{m_t^2}\right)^2 g(x)dx \\ \nonumber
& +  \frac{g^2}{64\pi}\frac{m_t^3}{M_W^2}\left(1-\frac{M_W^2}{m_t^2}\right)^2~.
\end{align}
The first term in Eq. (\ref{eqn:GammaGILongW}) contains all of the $d$ and $\mu$ dependence, while the second term is completely independent of $d$ and $\mu$. As a check on the correctness of Eq. (\ref{eqn:GammaGILongW}), note that for $d=1$, the Unhiggs model is equivalent to the SM and therefore the decay rates in the Unhiggs model and in the SM should be equal. This condition holds true by virtue of the fact that $g(x) = 0$ at $d=1$, and thus the first term in Eq. (\ref{eqn:GammaGILongW}) is also zero at $d=1$. The only non-zero part of the decay rate is the second term, which is exactly the SM result. As a further check of this result, the calculation was done in a general gauge, with $\xi$ left arbitrary. The calculation is considerably more involved, but the final gauge invariant decay rate remains, as it must, equal to the decay rate found in Eq. (\ref{eqn:GammaGILongW}).

\section{Comparison with Data}
\label{sec:data}
\setcounter{equation}{0}

\begin{figure}[h!]
\centering
\includegraphics[height = .4\textheight]{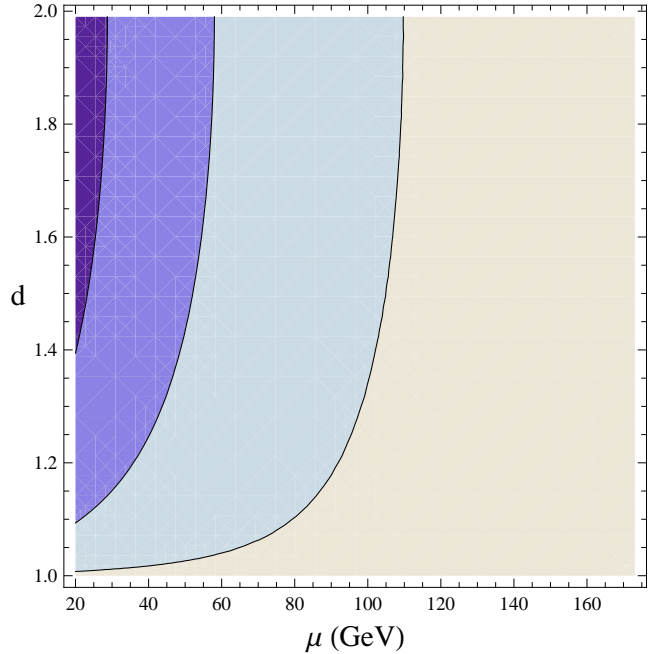}
\caption{Contour plot of $\mathcal{F}_0$ as a function of $\mu$ and $d$. The small dark region is the parameter space ruled out by current CDF data at a 90\% CL while the next darkest region is ruled out by CDF data at a 68\% CL. The next darkest (light blue) region is the additional parameter space affected by expected LHC data, also at a 68\% CL.}
\label{fig:bounds}
\end{figure}

To compare with current CDF data and future LHC data, we must calculate the value of $\mathcal{F}_0$, which is defined as the fraction of the top decays with a longitudinally produced $W$ boson: 
\beq
\mathcal{F}_0 \equiv \frac{\Gamma(t\rightarrow W^+_L b)}{\Gamma(t\rightarrow W^+_L b)+\Gamma(t\rightarrow W^+_T b)}~.
\eeq
with $\Gamma(t\rightarrow W^+_L b)$ given in Eq. (\ref{eqn:GammaGILongW}) and $\Gamma(t\rightarrow W^+_T b)$ given in Eq. (\ref{eqn:GammaTransW}). The current top quark data from CDF yields the following value for $\mathcal{F}_0$ \cite{F0CDF}:
\begin{align}
\label{eqn:F0CDF}
\mathcal{F}_0 = .66 \pm .16~.
\end{align}
The LHC promises to make a more accurate determination of this quantity. With an integrated luminosity of 10 fb$^{-1}$, the accuracy of the $\mathcal{F}_0$ measurement at the LHC should be at a level of about $\pm .015$ \cite{F0LHC}. The Standard Model tree level calculation for this quantity is $\mathcal{F}_0 = .699$, which is clearly within the allowed region in Eq. (\ref{eqn:F0CDF}). The value of $\mathcal{F}_0$ in the Unhiggs model will depend on the values of $\mu$ and $d$. Because the first term in Eq. (\ref{eqn:GammaGILongW}) is always greater than or equal to zero, and the second term is a constant equal to the SM result, the value of $\mathcal{F}_0 $ in the Unhiggs model will always be greater than or equal to the SM value for any values of $\mu$ and $d$. From Eq. (\ref{eqn:F0CDF}), we see that the Unhiggs model is thus ruled out at the 68\% level for $\mathcal{F}_0 > .82$. Figure \ref{fig:bounds} shows a contour plot of the $\mu$-$d$ parameter space that is constrained by CDF, and also the expected constraints on the parameter space due to the LHC.  In this plot, we assume the central value of $\mathcal{F}_0$ at the LHC to be the Standard Model value of 0.699,  not the current CDF value of 0.66. From Figure \ref{fig:bounds}, we see that the LHC greatly improves the constraints on the Unhiggs model for values of the Unhiggs threshold $\mu$ near $100$ GeV. For example, given a value of $\mu = 100$ GeV, current data says nothing about the value of $d$, but the LHC should constrain $d \lesssim 1.3$ at 68\%. However, for $\mu \gtrsim 110$ GeV, top decay analysis yields very little information about the value of the scaling dimension, even with expected LHC data. Of course, experimental cuts that restrict the longitudinal component to be close to the $W$ mass will further reduce the experimental sensitivity.

\section{Conclusions}
In this paper, we calculated the decay rate of the top quark in the Unhiggs model and found that, due to the unparticle nature of the Goldstone bosons, it is different than the top decay rate in the SM when the decay includes an emission of a longitudinal $W$ boson. We then compared this result to measurements from CDF, which constrain the fraction of top decays which contain a longitudinal $W$. This allowed us to rule out some regions of Unhiggs parameter space. The region of parameter space most affected by the CDF constraint is the region of high scaling dimension and low threshold mass (compared to the weak scale). Including the higher expected accuracy of top decay measurements at the LHC, we found that the LHC will be able extend the constrained region to intermediate values of $d$ and $\mu$. In future work, we hope to extend the study of Unhiggs phenomenology by directly calculating the Unhiggs production cross section and its various decay rates at the LHC.

\section*{Acknowledgements}
We thank Robin Erbacher, Jamison Galloway and Damien Martin for useful discussions and comments. We also thank Matthew Strassler for suggesting this calculation.

\end{document}